%% file: SynthesizingUpdateScheduleBMC.tex
\documentclass[submission,copyright,creativecommons]{eptcs}
\providecommand{\event}{FROM 2026} % Name of the event you are submitting to

\usepackage{iftex}

\usepackage[utf8]{inputenc}
\usepackage[T1]{fontenc}
\usepackage{graphicx}
\usepackage{amsmath}
\usepackage{amssymb}
\usepackage{todonotes}
\usepackage{url}
\usepackage{xspace}
\usepackage{array}
\usepackage{mathrsfs}
\usepackage{atbegshi}
\usepackage{minibox}
\usepackage{subcaption}
\usepackage{autobreak}
\usepackage{algorithm}
\usepackage{algpseudocode}
\usepackage{cite}
\usepackage{microtype}
\usepackage{csquotes}
\input{setup/settings}

\input{setup/macros}

\ifpdf
\usepackage[strings]{underscore}         % Only needed if you use pdflatex.
\else
\usepackage{breakurl}           % Not needed if you use pdflatex only.
\fi

\begin{document}

\title{Synthesizing Update Schedules with Game-Based Extension of Bounded Model Checking}

%\author{Janis Kröger \and  Paul Kröger \and Martin Fränzle
%\institute{Carl von Ossietzky Universität Oldenburg, Germany}
%\email{\{janis.kroeger,p.kroeger,martin.fraenzle\}@uni-oldenburg.de}
%}

\author{
	Janis Kröger \qquad Paul Kröger \qquad Martin Fränzle
	\institute{Carl von Ossietzky Universität Oldenburg, Germany}
	\email{\{janis.kroeger,p.kroeger,martin.fraenzle\}@uni-oldenburg.de}
}

\def\titlerunning{Synthesizing Update Schedules with BMC}
\def\authorrunning{J. Kröger et al.}

\maketitle

\begin{abstract}
Ensuring safe software updates in safety-critical systems without interrupting
    operation and without provisioning and activating cold spare hardware poses
    a fundamental challenge due to the conflict between system availability and
    update execution. In this paper, we present a bounded SMT encoding for synthesizing fixed global-time update schedules for timed-games with linear update automata and a fixed number of update transitions. %
    We model the interaction between the system and the update
    as a two-player timed game. Our key contribution is the synthesis of global
    time points that define a fixed update schedule which guarantees safe and
    complete deployment of the update independently of the autonomous system
    behavior. To this end, we reduce the scheduling problem to a reachability
    and safety objective and encode it as a quantified SMT problem. We
    demonstrate it on an example system of a trajectory planner for autonomous
    driving, showing that the synthesized schedule ensures safe deployment under
    all admissible executions.
\end{abstract}

\input{sections/sectintroduction}

\input{sections/sectbasics}

\input{sections/examplesystem}

\input{sections/sectconcept}

\input{sections/sectrelatedwork}

\input{sections/sectconclusion}

\bibliographystyle{eptcs}
\bibliography{references/bibliography}

\end{document}

%% file: setup/settings.tex
\newcommand{\secref}[1]{Section~\ref{#1}}

\newcommand{\figref}[1]{Figure~\ref{#1}}

\newcommand{\eqnref}[1]{Equation~\ref{#1}}

\newcolumntype{P}[1]{>{\centering\arraybackslash}p{#1}}

\newcolumntype{R}[1]{>{\raggedleft\arraybackslash}p{#1}}

\newcounter{speclines}

%% file: setup/macros.tex
\newcommand{\ie}{{{i.e.\/,}}}
\newcommand{\eg}{{{e.g.\/,}}}

%% file: sections/sectintroduction.tex
\section{Introduction}
\label{sec:introduction}

In the age of digitized infrastructures and cyber-physical systems (CPS),
software-heavy systems and software-defined systems (SDS) are becoming
increasingly important. Modern SDS require continuous updates in order to meet
the current safety standards and performance requirements, whether in
safety-critical infrastructures, production lines, or the automotive industry.
This trend towards updates will increase even further in the future
\cite{guissouma2018empirical}. At the same time, there is a growing demand that
these systems should never be completely shut down, as any interruption in such
applications has costly consequences in terms of production downtime or loss of
use.

One potential  approach to tackle this problem is to perform updates during
operation. However, it is important to mention that not all parts of a system
can be updated at any time: Only components that are inactive and remain
inactive or are safely switched off at the time of the update can be updated.
This yields a fundamental conflict of objectives: on the one hand, the
functionality of the system should be guaranteed without restriction. On the
other hand, parts of the system must be turned off for updates in order to be
able to install them safely. From a safety-critical perspective, points in time
at which the individual update steps can be carried out safely.

To avoid hindering the system functions, the update installation must
follow a schedule that ensures that the part of the (sub-)system that is being
updated is not in use during the allocated update time. Otherwise, the system
could enter an unpredictable state. The update installation schedule aims to
update the subsystem while maintaining the overall system functionality. 

One possible solution is to take advantage of times when the system parts to be
updated are not in use anyway. These time windows, often in the form of interval
bounds within the system reacts, can be used to perform updates even if the
exact reaction time within the interval remains variable. However, due to the
varying system behavior caused by environmental influences and internal
dynamics, it is not sufficient to simply  determine an update time. Instead, a
strategy or a robust update schedule is required that works reliably even in the
worst-case scenario.  To analyze and resolve the conflict between usage and
execution of updates, we propose a game-theoretical approach.  Game theory in
general offers a rigorous framework for representing and analyzing strategic
interactions between rational decision-makers, called players
\cite{maschler2020game}. The conflict situation is modeled as a game of these
players.

In this paper we want to show an approach to synthesize a schedule in the form of
robust plan for updating a system in operation using a game-based extension of
bounded model checking (BMC). Our approach using the BMC builds upon the
game-theoretic approach for scheduling updates described by Kröger
and Fränzle \cite{kroger2023updates}.

In our approach, we model the conflict situation of the update procedure as a
timed game, more precisely as a two-player timed game between
\begin{enumerate}
    \item player $U$ which models the update process and whose actions
        are fully controllable, and
    \item player $S$ which models the system and its environment and whose
        actions are completely uncontrollable.
\end{enumerate}

For this modeling, we exploit the structured behavior in form 
operating modes of the system, as well as the partitioning of the update process into
different phases\cite{rakow2022roles}. For the scheduling of the update, a
sequence of global time points for initiating the individual update actions is
synthesized, which forms an update plan which is robust against arbitrary system
behavior. 
If updates are applied to individual operating modes only as described
in~\cite{kroger2025ensuring}, the original problem can be reduced to solving a
reachability-safety problem aiming at reaching a state where the update is
completed while updating an active mode is avoided. This enables systematic
analysis and synthesis within the game-based framework.

The paper is structured as follows. \secref{sec:basics} provides an overview and
the necessary game theoretic foundations for our approach. For illustration of the approach, an example is introduced in
\secref{sec:examplesystem}. \secref{sec:model} presents our approach, including
the synthesis of a robust update plan. \secref{sec:relatedwork} discusses
related work, while \secref{sec:conclusion} provides a conclusion.

%% file: sections/sectbasics.tex
\section{Timed Games} \label{sec:basics}

Timed games extend game graphs \cite{Bloem2018} by adding real-valued clocks
that evolve over time, yielding \emph{timed game
automata}. In this setting, each player is modeled by an individual \emph{timed automaton}
\cite{alur1994theory}. The overall timed game consists of a parallel
composition of these automata in which the transitions of player $U$ are
considered to be fully controllable while player $S$' transitions are completely
uncontrollable, which requires a distinction between these transitions during
the robust plan synthesis.

\subsection{Timed Automaton}

A \emph{timed automaton} is a tuple
\begin{equation*}
	A = (L, L^0, C, \Sigma, E, I),
\end{equation*} 
where:
\begin{itemize}
	\item $L$ is a finite set of \emph{locations},
	\item $L^0 \subseteq L$ is the set of initial locations,
	\item $C$ is a finite set of real-valued \emph{clock variables} $c \in
        \mathbb{R}_{\geq 0}$,
	\item $\Sigma$ is set of actions,
	\item $E \subseteq L \times \Phi(C) \times \Sigma \times 2^C \times L$ is a
        finite set of transitions $(\ell,\phi,a,X,\ell')$, with source location
        $\ell$, guard $\phi$, action~$a$, reset set $X$, and target location
        $\ell'$,
	\item $I : L \to \Phi(C)$ assigns an \emph{invariant} to each location.
\end{itemize}
Here $\Phi(C)$ denotes  the set of clock constraints over $C$ given by conjunctions
of atoms $x \sim k$ ($x\in C$, $k\in\mathbb{Q}$, $\sim\in\{<,\le,=,\ge,>\}$).

\paragraph*{Semantics of timed automata.} Given some timed automaton $A = (L,
L^0, C, \Sigma, E, I)$, let $v : C \to \mathbb{R}_{\geq 0}$ be
a clock valuation function. A pair $(\ell, v)$ with $\ell \in L$ is a
\emph{state} of $A$. The semantics of $A$
is given by a set of runs where
a \emph{run} of a timed automaton $A$ is an infinite alternating sequence of delays and discrete transitions
\begin{equation*}
(\ell^0,v^0)\xrightarrow{t^0}(\ell^0,v^0+t^0)\xrightarrow{a^0}(\ell^1,v^1)\xrightarrow{t^1}\cdots
\end{equation*}
where $t^i\in\mathbb{R}_{\ge0}$ and $a^i\in\Sigma$.  
For each discrete transition $(\ell,v)\xrightarrow{a}(\ell',v')$
there exists a transition $(\ell,\phi,a,X,\ell')\in E$ such that
$v\models\phi$ and $v'=v[X:=0]$ where $v[X:=0](c) = 0$ iff $c \in X$, and
$v[X:=0] = v(c)$ otherwise, and $v' \models I(\ell')$.
Each time transition $(\ell,v)\xrightarrow{t}(\ell,v+t)$ requires $v+d\models
I(\ell)$ for all $d\in[0,t]$ and where $(v+t)(c) = v(c) + t$.

To track elapsed time, we introduce a \emph{global clock} $c_g\notin C$ with
$v^0(c_g)=0$ that is never reset. Hence $v(c_g)$ equals the total elapsed time
of the run.

\subsection{Parallel Composition} \label{subsec:parallelcomp}

Let $A_i=(L_i,L_i^0,C_i,\Sigma_i,E_i,I_i)$ for $i\in\{U,S\}$ with disjoint clock
sets $C_S\cap C_U=\emptyset$. Furthermore, let $\bot_i$ be a special action
indicating that no action is chosen, and let $E_{\bot,i} = \{(\ell_i,
\mathrm{true}, \bot_i, \emptyset, \ell_i) \mid \ell_i \in L_i \}$ be the set of
\emph{stutter transitions} that
does not change the state of $A_i$ and which is taken whenever action
$\bot_i$ is chosen.
The parallel composition $A_{\parallel}=A_S\parallel
A_U$ is a timed automaton
$A_\parallel =(L_\parallel, L^{0}_{\parallel}, C_{\parallel},
\Sigma_{\parallel}, E_{\parallel}, I_{\parallel})$ with 
\begin{equation*}
    L_\parallel = L_S\times L_U, \;\; L_{\parallel}^{0} = (L_S^0,L_U^0),\;\;
    C_{\parallel} = C_S\cup C_U,\;\; \Sigma_{\parallel} = (\Sigma_S \cup
    \{\bot_{S}\})\times (\Sigma_U \cup \{\bot_U\})
\end{equation*}
and with invariant
\begin{equation*}
    I_\parallel((\ell_S,\ell_U))=I_S(\ell_S)\land I_U(\ell_U).
\end{equation*}
for $\ell_i \in L_i$ and with
\begin{equation*}
    E_{\parallel} \subseteq L_\parallel \times \Phi(C_\parallel)
    \times \Sigma_\parallel \times 2^{C_\parallel} \times L_\parallel
\end{equation*}
s.t.\ for each $(\ell, \phi, a, X, \ell') \in E_{\parallel}$ there are
transitions $(\ell_i, \phi_i, a_i, X_i, \ell_i') \in E_i \cup E_{\bot,i}$ with $\ell =
(\ell_S, \ell_U)$, $\phi = \phi_S \land \phi_U$, $X = X_S \cup X_U$, $a =
(a_{S}, a_{U})$ and $\ell' = (\ell_S', \ell_U')$.

\subsection{Games, Winning Conditions and Strategies}\label{subsec:GWCS}

In a game-theoretic interpretation of $A_\parallel$, player $U$ is modeled by
$A_U$, and player $S$ is modeled, by $A_S$. A run -- or \emph{play} in the
game-theoretic context -- of $A_\parallel$ is said to be
\emph{winning} for player $U$ if it satisfies some condition. The goal is to
find a sequence of actions for player $U$ such that for all possible actions of
player $S$, player $U$ wins.

\paragraph{Winning conditions.}

Let $T\subseteq L$ be a set of \emph{target} locations and $B\subseteq L$ a set
of \emph{bad} locations.  For a play $\rho=(\ell^0, v^0)(\ell^1, v^1)\cdots$, we
define:
\begin{align*}
\mathrm{Reach}(T)=\{\rho \mid \exists k\ge0:\ell^k\in T\},\\
\mathrm{Safe}(B)=\{\rho \mid \forall k\ge0:\ell^k\notin B\},\\
\mathrm{ReachSafe}(T,B)=\mathrm{Reach}(T)\cap\mathrm{Safe}(B).
\end{align*}
Thus player $U$ aims at eventually reaching a location in $T$ while never visiting any location in $B$.  

\paragraph{Robust plans.}

A robust plan is a fixed sequence of actions s.t.\ whatever actions player
$S$ chooses, player $U$ wins. Let $n$ be the number of transitions for
player~$U$ necessary to win the play. Then, in our setting of timed games, the
robust plan consists
of a finite sequence $\sigma = \langle t_{g_1}, \cdots, t_{g_n}\rangle$ of time
instances at which player~$U$ takes a transition instead of actions from a
dedicated action set $\Sigma_U$.
Given $T$ and $B$, the objective of player~$U$ is to enforce
$\mathrm{ReachSafe}(T,B)$ against all behaviors of player $S$ and its
environment.

\paragraph{Assumption of linear deterministic update automaton.}\label{assumption1} In this work, we consider an update automaton $A_U$ in which the update path from the initial location to the final location is linear and deterministic. For each synthesized global time point $t_{g_j}$, at most one controllable update transition is enabled. Hence, once $t_{g_1}, \cdots, t_{g_n}$ are fixed, the update behavior is
uniquely determined and cannot branch depending on future system behavior.

\subsection{Bounded Model Checking:} \label{subsec:bmc}

Bounded model checking (BMC) is a symbolic verification technique for analyzing
the behavior of transition systems within a finite execution bound. A system is
described by an initial condition and a transition relation, which are unrolled
up to a fixed bound $k$, resulting in a formula over the state variables at each
step.

The verification task is then reduced to a satisfiability problem: BMC checks
whether there exists an assignment to all variables of the unrolled system such
that the resulting formula is satisfied. By additionally encoding a property
$\varphi$ over the bounded execution, this corresponds to checking whether there
exists a run of length up to $k$ that satisfies the given property.
Thus, BMC transforms the analysis of system behaviors into a constraint-solving
problem, where satisfying assignments directly correspond to concrete executions
of the system within the given bound.

Formally, let a system be defined by an initial condition $Init(s^0)$ over the
initial state $s^0$, and a transition relation $Trans(s^i, s^{i+1})$ describing
valid successor states. A system evolution over $k$ steps is then characterized
by the formula
\begin{equation*}
Init(s^0) \;\wedge\; \bigwedge_{i=0}^{k-1} Trans(s^i, s^{i+1}).	
\end{equation*}

To check a property $\varphi$, bounded model checking encodes the existence of a
system evolution of length at most $k$ that satisfies the specification as a
satisfiability problem. This is achieved by constructing a formula of the form
\begin{equation*}
Init(s^0) \;\wedge\; \bigwedge_{i=0}^{k-1} Trans(s^i, s^{i+1}) \;\wedge\;
    \bigvee_{i=0}^{k} \varphi(s^k),	
\end{equation*}
If the resulting formula is satisfiable, the satisfying assignment
corresponds to a run of length at most $k$ that fulfills the property. By
increasing the bound $k$ iteratively, evolutions of increasing length can be
explored.

While BMC in this form searches for satisfying executions within a bounded horizon, it remains inherently incomplete for unbounded reasoning, as it only considers runs up to the given bound. Nevertheless, it provides a scalable and flexible framework that can be extended beyond verification, for example, to the synthesis of strategies by introducing decision variables and additional constraints into the underlying formula.

%% file: sections/examplesystem.tex
\section{Example System: Trajectory Planner}\label{sec:examplesystem}

To illustrate our approach, we consider a trajectory planner, which is a central
component of a highway pilot system responsible for generating feasible
trajectories for a vehicle. The system enables safe autonomous driving on
the highway under normal conditions and through construction zones.  While the
primary function of the trajectory planner is to compute trajectories based on
sensor inputs, in the context of this work we focus exclusively on its
\emph{mode-dependent behavior} and the conditions for \emph{mode transitions}.

The trajectory planner operates in four distinct modes:
\begin{itemize}
	\item \textit{Off (Off)}: No trajectory planning is performed, \eg{} during manual driving or when the system is inactive.
	\item \textit{Normal (N)}: Default operation on regular highway lanes under nominal conditions.
	\item \textit{Construction Zone (CZ)}: Specialized operation in construction zones, requiring adapted handling of lane markings and environmental constraints.
	\item \textit{Emergency (E)}: Safety mode activated in the presence of critical faults or unreliable input data.
\end{itemize}

Mode transitions are triggered by a combination of external requests and system conditions derived from other components, in particular a lane detection and an object detection. 
The transition structure can be summarized as follows.

Transitions between \textit{Off} and \textit{Normal} are exclusively triggered by an external \textit{Request}, which activates or deactivates the trajectory planner. The transition from \textit{Normal} to \textit{Construction Zone} is conditioned on consistent environmental information. The lane detection and object detection must indicate construction zone conditions. If these conditions are no longer satisfied, the system returns to \textit{Normal} mode. 
A transition to \textit{Emergency} mode may occur from any active mode (\textit{Normal} or \textit{Construction Zone}) if the input data from lane detection or object detection is faulty, incomplete, or unavailable. The \textit{Emergency} mode represents a fail-safe state in which normal operation is suspended. A change from \textit{Emergency} is only possible via an external \textit{Request}, which forces a transition to \textit{Off}.

From a timing perspective, mode transitions are not instantaneous. Each transition takes $5$--$15\,ms$. These timing constraints are relevant for ensuring safe coordination with other components and must be respected when reasoning about system behavior.
Since a \textit{Request} can occur at any time, the mode can change from \textit{Off} to \textit{Normal}, or from \textit{Normal}, \textit{CZ} or \textit{Emergency} to \textit{Off} at any time. The trajectory planner receives inputs periodically, every $30\,ms$. A decision based on this input  is made every $30\,ms$ as to whether it should change modes between \textit{Normal}, \textit{CZ} and \textit{Emergency}.

In our example, we now want to update the \textit{CZ} mode to improve performance. The update takes $10\,ms$ and is performed in a single step.

%% file: sections/sectconcept.tex
\section{Bounded Model Checking for Safe Updates with Game Based Extension}\label{sec:model}

We model the interaction between the running system and the update process as a
timed game.  The system with its environment is fully uncontrollable and act as
player~$S$, while the update is fully controllable (player~$U$). As we mentioned
in \secref{subsec:parallelcomp}, we model both players as individual timed
automaton that run in parallel.

The objective of the update is to safely update a specific system mode while the
system continues its operation.  This is formalized as a reachability safety
condition: the update must eventually reach its final location while the system
never enters the mode currently being updated.

To achieve this, we synthesize a schedule in form of a robust plan of update
actions using a game-based extension of bounded model checking.  Instead of
computing a fully reactive strategy as it is possible with UPPAAL
\cite{behrmann2006uppaal, behrmann2007uppaal}, we search for a robust update
plan that succeeds for all possible system behaviors.

For this robust plan, we identify global points in time at which the update can
perform a corresponding transition in the automaton.  These global time points
define when a transition in the update automaton $A_U$ must be taken to
guarantee safe completion of the update independently of the system behavior.

\subsection{Modeling of the Player}\label{subsec:modelingplayer}

We model both the update and the system as timed automata. 
We treat the modes of the system as location in the automaton.
The trajectory planner is modeled as a timed automaton whose behavior is fully characterized by its modes and the corresponding transition conditions, as described in \secref{sec:examplesystem}. The timing constraints are modeled accordingly as guards and invariants.
The resulting automaton for the system is 
\begin{equation*}
	A_S=(L_S,L_S^0,C_S,\Sigma_S,E_S,I_S).
\end{equation*}

To model the update, we consider the phases of an update from
\cite{rakow2022roles} as modes and map them accordingly to locations in the
automaton.  For simplicity, we treat the transmission, installation, and
clean-up phases as a combined phase \textit{Update}.  In addition to the timing
constraints, which are modeled as invariants and guards, the update automaton
has an additional guard at each transition for the global time ($t_g$) at which
this transition can be taken. This time is crucial for the synthesis.    The
result of the update is the automaton:
\begin{equation*}
	A_U=(L_U,L_U^0,C_U,\Sigma_U,E_U,I_U)
\end{equation*}

\begin{figure}[tb]
	\centering
	\begin{subfigure}{0.15\textwidth}
		\centering
		\includegraphics[width=\textwidth]{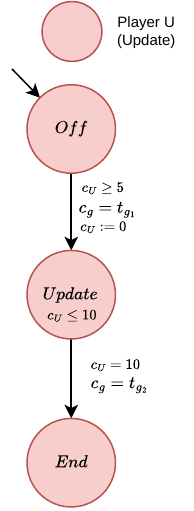}
	\end{subfigure}
	\hfil
	\begin{subfigure}{0.3\textwidth}
		\centering
		\includegraphics[width=\textwidth]{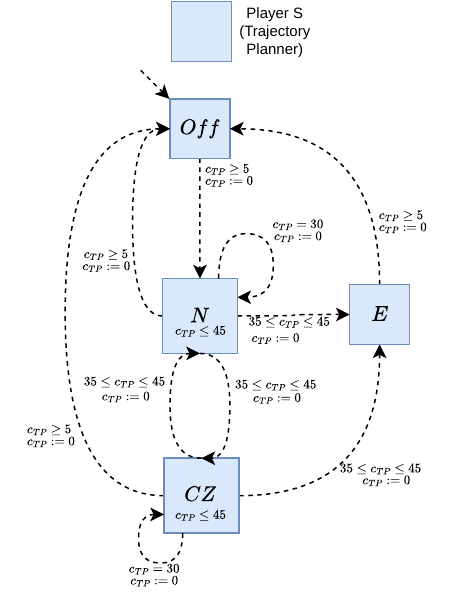}
	\end{subfigure}
	\caption{Timed automaton of the update (left side) and the system (right
    side).}
	\label{fig:playerautomaton}
\end{figure}

\figref{fig:playerautomaton} illustrates the two automata for our example. Dotted lines indicate uncontrollable transitions, and solid lines indicate controllable transitions.

Both automata run in parallel as described in \secref{subsec:parallelcomp}.  
The guards in the system automaton are determined by the transition times for a mode change ($5-15\,ms$) and the corresponding time interval between the periodic occurrence of the inputs ($30\,ms$).
The resulting game model describes the interaction between the update controller and the system environment.  

In principle, an update can be completed before the estimated time. However, since this cannot be guaranteed for all system behaviors, we use the maximum estimated update duration when synthesizing the robust plan  ($10\,ms$ in our example). Although this results in a more conservative schedule, it enables us to ensure the update is executed without conflicts. 
Consequently, if an update finishes earlier in practice, the synthesized strategy remains valid.

\paragraph{Winning Conditions.}

The goal of player~$U$ is to successfully complete the update while avoiding \emph{bad} locations.  
Let $End$ denote the final location of the update automaton and $CZ$ as the mode to be updated.  
We define the target set $T$ and the set of \emph{bad} locations $B$:

\begin{equation*}
	T = \{End\}\times L_S \quad, \qquad 	B = \{(Update,CZ)\}.
\end{equation*}

Player~$U$ must therefore satisfy the reach-safety objective
\begin{equation*}
\mathrm{ReachSafe}(T,B).
\end{equation*}

\subsection{Synthesis of Robust Plan via Bounded Model Checking}

We synthesize a robust plan by encoding the timed game as a satisfiability problem using a game-based extension in form of a quantifier alternation within bounded model checking (BMC).  
To this end, the behaviors of the involved automata are unrolled over a bounded number of steps $k$, resulting in a finite transition system that is encoded as an SMT formula over a set of variables.  
These variables include boolean location variables, real-valued clock variables, and necessary variables introduced for the game encoding.

\subsubsection{Global Time}

Since we aim to synthesize strategies based on absolute time points, we first introduce a global notion of time.  
For this purpose, we  introduce $k+1$ real-valued variables
\begin{equation*}
	c_g^0, \dots, c_g^k,
\end{equation*}
where $c_g^i$ represents the global clock at step $i$.

The evolution of the global clock is constrained by an initialization predicate and a transition relation.  
We first define the initial condition
\begin{equation*}
Init_{global}(c_g^0) \equiv (c_g^0 = 0),
\end{equation*}
which ensures that the global clock starts at zero.  
The progression of time is modeled by the transition predicate
\begin{equation*}
Trans_{global}(c_g^i, c_g^{i+1}) \equiv 
(c_g^{i+1} > c_g^i) \wedge (c_g^i \geq 0) \wedge (c_g^{i+1} \geq 0),
\end{equation*}
enforcing strictly monotonic time evolution.

The valid evolution of the global clock is denoted by \emph{Valid Run Clock} (VRC):
\begin{equation*}
VRC \equiv Init_{global}(c_g^0) \wedge \bigwedge_{i=0}^{k-1} Trans_{global}(c_g^i, c_g^{i+1}).
\end{equation*}

\subsubsection{Encoding of System and Update Runs}

We now encode the behaviors of the system and the update automata.
Let $\vec v^i_S$ denote the set of all variables of the system automaton at step $i$, and $\vec v^i_U$ for the update automaton.  
For each automaton, we define an initial condition $Init(\vec v^0)$ and a transition predicate $Trans(\vec v^i,\vec v^{i+1})$ describing valid successor states.

The \textit{Valid Run of the System} (VRS) automaton is given by
\begin{equation*}
VRS \equiv Init_S(\vec v^0_S) \wedge \bigwedge_{i=0}^{k-1} Trans_S(\vec v^i_S,\vec v^{i+1}_S),
\end{equation*}
which encodes all valid behaviors of the system.

For the update automaton, we define the \emph{Valid Run of the Update} (VRU) as
\begin{equation*}
VRU \equiv
Init_U(\vec v^0_U)
\land
\bigwedge_{i=0}^{k-1} Trans_U(\vec v^i_U,\vec v^{i+1}_U)
\land
\bigvee_{i=0}^{k} (End_U^i).
\end{equation*}
Thus, $VRU$ combines the validity of the update run, the reachability objective, and the encoded safety objective.
Therefore, the transition predicate $Trans_U$ of the update includes a constraint to avoid the defined \emph{bad} locations, while the predicate $End_U$ encodes the reachability condition, \ie{} that a designated target location is eventually reached within $k$ steps.

\paragraph{Location Encoding.}
For each location $\ell \in L$ we introduce a set of $k+1$ boolean variables. $\ell^i$ indicating that the automaton is in location $\ell$ at step $i$.  
Exactly one location variable must be true at each step.

\paragraph{Clock Encoding.}
For every clock $c\in C$ we introduce $k+1$ real-valued variables
$c^0,\dots,c^k$
representing the clock valuation at each step.  
Time progression between steps depends on the progress of the global clock:
\begin{equation*}
c^{i+1} = c^i + (c_g^{i+1} - c_g^i).
\end{equation*}
The following applies when resetting a clock as a result of a transition: 
\begin{equation*}
	c^{i+1} = 0.
\end{equation*}
Within the \textit{Trans} predicate, we distinguish between so-called \textit{stutter} transitions  ($\bot$) and \textit{real} transitions. In so-called \textit{stutter transitions}, time simply elapses without a subsequent change of location. In a \textit{real} transition, a change of location follows the elapsed time. 

\paragraph{Global Time Points.}
In the $VRU$, there is an additional guard at the real transitions that must be satisfied. Every real transition is associated with a specific global time (see \figref{fig:playerautomaton}). Therefore, it can only be taken if the time corresponds to the global clock, \eg{} $t_{g_1} = c_g^i$.

To ensure that the global time points $t_{g_1}, \dots, t_{g_n}$ are properly reflected in the behavior of player~$S$, we restrict the considered evolution of the global clock $c_g$ to those in which these time points are actually attained.
To this end, we introduce the constraint $tg_{\mathit{cons}}$, which enforces that each $t_{g_j}$ coincides with the value of $c_g$ at some step of the execution. Formally, for each $j \in {1, \dots, n}$, we require that $t_{g_j}$ matches the value of $c_g$ at some point over all steps:
\begin{equation*}
	tg_{\mathit{cons}} \equiv \bigwedge_{j=1}^{n} \left( \bigvee_{i=0}^{k} t_{g_j} = c_g^i \right).
\end{equation*}

For illustration, in \eqnref{eq:transalterexample}, an excerpt of the \textit{Trans} predicate for our example system, specifically for the transition from \textit{Off} to \textit{Normal}, is encoded. In \eqnref{eq:transegoexample}, an excerpt of the \textit{Trans} predicate for the update, from \textit{Off} to \textit{Update}, including the safety properties, is encoded.   

\noindent
\begin{minipage}{\textwidth}
\begin{minipage}[t]{0.44\textwidth}
	\vspace{0pt}
	\begin{equation}
		\begin{aligned}
		Tran&s_{TP}(\vec v^{i}_{TP}, \vec v^{i+1}_{TP}) = \Bigg(\cdots \vee \\
		&\Big( Off^{i} \wedge \neg Off^{i+1}  \\ 
		&\quad \wedge \neg N^{i} \wedge N^{i+1}\\ 
		&\quad \wedge \neg E^{i} \wedge \neg E^{i+1} \\ 
		&\quad \wedge \neg CZ^{i} \wedge \neg CZ^{i+1} \\ 
		&\quad \wedge c_{TP}^{i} + (c_{g}^{i+1} - c_{g}^{i}) \geq 5\\
		&\quad \wedge c_{TP}^{i+1} = 0 \\ 
		&\quad \wedge z^{i+1} = z^{i} + 1 
		\Big) \vee \cdots \Bigg) 
\end{aligned}
\label{eq:transalterexample}
\end{equation}
\end{minipage}
\hfil%
\begin{minipage}[t]{0.47\textwidth}
\vspace{-12pt}
\begin{equation}
\begin{aligned}	
		Tran&s_{U}(\vec v^{i}_{U}, \vec v^{i+1}_{U}) = \Bigg( \neg (Update^{i} \wedge CZ^{i}) \\ &\quad \wedge \neg (Update^{i+1} \wedge CZ^{i+1}) \\ 
		&\quad \wedge \Big(Off^{i} \wedge \neg Off^{i+1} \\
		&\quad \wedge \neg Update^{i} \wedge  Update^{i+1} \\ 
		&\quad \wedge \neg End^{i} \wedge \neg End^{i+1} \\ 
		&\quad \wedge c_{U}^{i} + (c_{g}^{i+1}-c_{g}^{i}) \geq 5\\
		&\quad \wedge c_{U}^{i+1} = 0 \\ 
		&\quad \wedge c_{g}^{i+1} = t_{g_1} \Big) \vee \cdots \Bigg)
\end{aligned}
\label{eq:transegoexample}
\end{equation}
\end{minipage}
\label{eq:twoformulas}
\end{minipage}

\begin{figure}[tb]
	\centering
	\begin{subfigure}{0.49\textwidth}
		\centering
		\includegraphics[width=\textwidth]{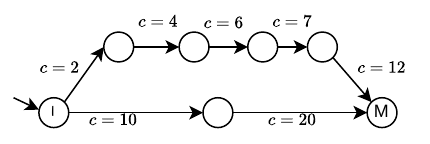}
	\end{subfigure}
	\hfil
	\begin{subfigure}{0.49\textwidth}
		\centering
		\includegraphics[width=\textwidth]{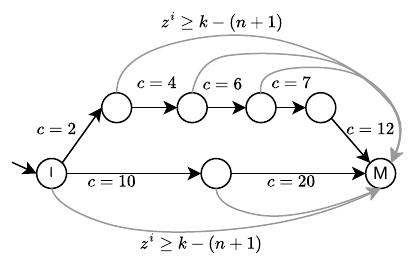}
	\end{subfigure}
	\caption{Timed automaton without \emph{z-based transition} (left side) and with grey \emph{z-based transition} (right
		side). For better readability, the guards have been omitted at some transitions.}
	\label{fig:zbasedtransition}
\end{figure}

\paragraph{z-Based Transitions.}
A key problem in BMC is that the step bound may be too small. Consequently, the method may only consider runs that are reachable within this bound. Longer runs with more transitions that lead to the target location in less real time are not considered. As a result, the synthesized time points may seem correct within the chosen step size, but may in fact be unsafe, since the system could potentially reach the target state earlier via an ignored path. 
Let us consider the timed automaton on the left side in \figref{fig:zbasedtransition}, where mode $M$ is to be updated, \ie{} not taken, while the update is being performed, and an update automaton as shown in \figref{fig:playerautomaton}.
The BMC algorithm can provide a solution for the global time points after just two steps, \eg{} $t_{g_1} = 5$ and $t_{g_2} = 15$.
However, this solution would only be safe if the system follows the bottom path, which can be reached after two steps. For the top path, the system requires more steps but less time and would reach mode $M$ and thus a \emph{bad} location within 12 time units. To address this problem, we introduce so-called \emph{z-based transition} for player $S$ (grey transition at the automaton on the right side of in \figref{fig:zbasedtransition}).

The purpose of the z-based transition is to conservatively consider system paths in which a budget of transitions for the system is determined by the number of discrete system transitions that have been taken, depending on the depth $k$ of the BMC and the number of fixed transitions $n$ of the update automaton.

Due to the number of $n$ transitions and the constraint that the value of the global clock $c_g$ must correspond to a synthesized time in these steps, $n$ transitions are determined based on the possible values depending on the behavior of the update automaton.
Since the initial values of the variables are fixed for both players, there is a possible free budget of transitions, resulting in a possible variable assignment with a depth of $k-(n+1)$.   

Once the system has used up the budget, a \emph{z-based transition} leads directly to a designated \emph{bad} location, which must be avoided during the update process. This prevents the synthesized schedule from being misclassified as safe simply because a path that is faster in terms of time but has more transitions in the required transitions is excluded by the boundary.

Since the available budget for a run depends on the system transitions that have already taken, we introduce a set $k+1$ of counting variables $z^0, \dots, z^k$.
The counter increments whenever player~$S$ performs a real transition, \ie{} $z^{i+1} = z^i + 1$, and remains unchanged for stutter transitions, \ie{} $z^{i+1} = z^i$.
A \emph{z-based transition} at step $i$ is enabled if the following guard holds:
\begin{equation*}
	z^i \geq k - (n + 1),
\end{equation*}
where $n$ denotes the number of fixed transition of the update automaton where $t_g$ that are synthesized. 
These transitions ensure that the BMC algorithm considers a sufficient number of steps, taking into account all necessary system behavior.

\subsubsection{Game Semantics Encoding}
Based on the previously defined formulas $VRS$, $VRU$, $VRC$, and $tg_{cons}$, we now construct the overall game encoding as a quantified formula.

Let $\vec v_{U}$ denote all the variables controlled by player~$U$ except from the synthesized global time points $t_g$. $\vec v_{S}$ denote all variables of the system and $\vec c_{g}$ the variables representing the global clock variables. All variables are defined over the bounded horizon of $k$ steps.

Since we want to synthesize a robust plan that specifies when the transitions of the update automaton can be taken, 
we consider the following implication:

\begin{equation*}
(VRS \wedge VRC \wedge tg_{cons}) \Rightarrow VRU,
\end{equation*}
which states that for a valid system evolution and global clock evolution that realize the chosen time points, the update run must satisfy the defined reach-safety objectives we defined in \secref{subsec:modelingplayer} and encoded in $VRU$.

The synthesis objective is to ensure that the update succeeds for all admissible behaviors of the environment.  This is captured by the following quantified formula:
\begin{equation*}
\forall \vec v_{S}, \vec v_{c_g}
\;\exists \vec v_{U}
:
\neg (VRS \wedge VRC \wedge tg_{cons}) \vee VRU.
\end{equation*}

The quantifier prefix represents the desired correctness property: the environment actions are universally quantified and can thus evolve arbitrarily, considering valid behavior of the system ($VRS$) and valid clock evolution ($VRC$). The commitment to a schedule is represented solely by the existential selection of global time points.

This correctness property could now be refined into a formalization of a reactive game by reshuffling its quantifier prefix into an alternation of actions, reflected by an alternation of universal and existential quantifiers as usual in two-player zero-sum games with boolean rewards \cite{maschler2020game}. 
To synthesize a robust plan, we however strengthen it further. Starting from the game formalization, we first strengthen the formula by moving the existential quantifiers reflecting updater actions in front. As the update sequence is deterministic, it suffices to quantify existentially over the update time points, which are the only choices the update process has. This implies that a quantifier prefix of the form $\exists t_{g_1}, \dots, t_{g_n}$ fixes all choices of the update process, and that appending a universal quantifier block resolving all system choices leads to a formalization of robust plan generation. We thus arrive at the following formula for robust plan generation: 
\begin{equation*}
\exists t_{g_1}, \dots, t_{g_n}
\;\forall \vec v_{S}, \vec v_{c_g}
\;\exists \vec v_{U}:
\neg (VRS \wedge VRC \wedge tg_{cons}) \vee VRU,
\end{equation*}
where the time points $t_{g_1},\cdots,t_{g_n}$ represent the scheduled times of the individual update actions.
A satisfying assignment to the synthesized global time points $t_{g_1},\cdots,t_{g_n}$ yields a robust update plan that guarantees successful completion of the update independently of the system behavior within the considered bound.
Due to the mentioned assumption of a linear deterministic update automaton, these time points uniquely determine the execution of the update. Therefore, the existential variables $v_U$, merely reflect the resulting execution and do not represent additional reactive choices based on future system behavior.

To obtain the plausible values for $t_{g_1}, \cdots, t_{g_n}$, 
we must restrict the time points $t_{g_1}, \cdots, t_{g_n}$ by suitable constraints, which we refer to as \emph{GlobalConst} in the following:
\begin{equation*}
		GlobalConst = (t_{g_1} < t_{g_2} < \cdots < t_{g_n}) 
		\wedge  (t_{g_1} > 0)
		\wedge  \cdots
		\wedge  (t_{g_n} > 0) \,.
\end{equation*}

Since these constraints relate existential values, we conjoin this \emph{GlobalConst} with the overall formula.
This yields the complete formula:

\begin{equation*}
	\begin{split}
		\phi(k,n) = \, &\exists\, t_{g_1}, \cdots, t_{g_n}:  \\	
		&\qquad GlobalConst \wedge \forall\,\vec v_{S},  \vec v_{c_g} \; \exists \vec v_{U}:\quad \\
		&\qquad\qquad\qquad\lnot(VRS \wedge VRC \wedge tg_{cons}) \vee VRU \,.
	\end{split}
\end{equation*}

\paragraph{Soundness of the Encoding.} Let $A_S$ be the system automaton and $A_U$ a deterministic linear update automaton satisfying our assumption in  \secref{assumption1}. If the SMT formula $\phi(k,n)$ is satisfiable and yields global time points  $t_{g_1}, \cdots, t_{g_n}$, then for all valid system executions within $k$ steps and satisfying $tg_{cons}$, the induced update execution reaches the $End$ location without visiting a bad location within $k$ steps. 

The quantified formula considers all valid system behavior and global clock evolutions within k steps that satisfy the conditions $VRS$, $VRC$, and $tg_{cons}$.
Based on our assumption, the synthesized global time points uniquely determine the execution of the update. 
Therefore, if the formula $\varphi(k,n)$ is satisfiable, then the reach-safety objective encoded in $VRU$ holds for all admissible system behaviors within the chosen bound.

If the SMT formula $\varphi(k,n)$ is unsatisfiable, then  no robust plan exists for the chosen number of $k$ steps together with the number of transitions $n$ in the update automaton $A_U$. This does not imply that no robust plan exists in general in the unbounded timed game. 

The used \emph{z-based transitions} in the system over-approximate the system behavior, which may require more discrete transitions than are available within the current bounds to reach an unsafe location. Therefore, the inclusion of these transitions makes satisfiability more conservative: if the robust plan is safe against the enlarged behavior of the system, then it is also safe against the original system behavior. Thus, the \emph{z-based transitions} provide a conservative over-approximation of the original system behavior.

\paragraph{Algorithm.}
The overall synthesis procedure is summarized in Algorithm~\ref{alg:gamebmcfinal} for our example system. 
It provides the game-based extension by introducing alternating quantifiers to capture the interaction between the system and its environment, and synthesizes global time points that define a robust update plan. 
If no upper bound is specified, the procedure is semi-decision-like: satisfiable instances may be found by increasing $k$, while unsatisfiability in the unbounded game cannot be established. For the example implementation, we use a user-specified maximum bound, $k_{max}$.
If this limit is reached without any satisfiable instances, we report \enquote{\texttt{unsat} up to $k_{max}$}.
\begin{algorithm}[tb]
	\caption{Game-Based BMC with Global Time Points}
	\label{alg:gamebmcfinal}
	\begin{algorithmic}[1]
		\Require Initial unrolling depth $k_{min}$, max depth $k_{max}$, fixed transition  number $n$ of $A_U$
		\Ensure Global time points $t_{g_1}, \dots, t_{g_n}$ or \texttt{unsat/unknown} up to bound
	
		\For{$k = k_{min}$ \textbf{to} $k_{max}$}
		\State \quad Instantiation of solver $S$
		\State \quad construct $\phi(k,n)$
		\State \quad $S.add(\phi(k,n))$
		\If{$S.check() = sat$}
		\State \Return $model(t_{g_1}, \dots, t_{g_n})$
		\Else
		\State  \texttt{unsat} for $k$ steps 
		\EndIf
		\EndFor
		\State \Return \texttt{unsat} up to $k_{max}$
	\end{algorithmic}
\end{algorithm}

\subsubsection{Result of the Example System}
To demonstrate our synthesis and to solve the constraints encoded by the overall formula, we use the SMT-solver Z3 \cite{de2008z3}. 
The solver requires $k = 6$ steps and outputs the solutions $t_{g_1} = 5$ and $t_{g_2} = 15$. For these global time points, it can be guaranteed that the transitions in the update automaton can be performed, regardless of what the system has done in the meantime. This means that the update can be started at time $t_{g_1} = 5$ and finished at time $t_{g_2} = 15$.  
For all $k<6$ we get the report of \enquote{\texttt{unsat} for $k$ steps}. 
Another possible interpretation of the synthesized time points is to consider them in relation to the transitions of the system. In this context, each time point at which the system transitions to its initial state, it serves as a new reference point for the synthesized time points as long as the update process has not started. The update process is initiated only with the first transition. This means for our example that, every time the
\textit{Trajectory Planner} changes to the \textit{Off} mode, the update can be started after $5\,ms$ by a transition of the update.

\subsection{Limitations and Discussion}
In our example the algorithm cannot generate a robust plan for updates longer than $35\,ms$. This stems from the non-determinism of the \textit{Trajectory Planner} with respect to its mode transitions. Beyond $35\,ms$, the location of the \textit{Trajectory Planner} within the automaton cannot be determined reliably. Moreover, there is no guarantee of a time after which the planner will re-enter an updatable mode. 

In such scenarios, a combination of a robust plan with a so called reactive strategy, which we can synthesize, \eg{} using UPPAAL TIGA \cite{behrmann2007uppaal}, can be helpful.
This combination means that we synthesize global points in time at which the update can be performed its transitions up to a certain point. For subsequent update transitions, we synthesize a reactive strategy that specifies how the update should behave in relation to system behavior to satisfy the winning condition.

The use of absolute time is crucial for the synthesis of robust plans.  
In a relative-time semantics, both players propose local delays, and the progression of time depends on their interaction. Due to nondeterministic delays, unbounded residence times in locations, and the universal quantification over the environment behaviors, the same BMC step may correspond to different absolute time points across executions.
As a consequence, it is not possible to determine the exact system state at a given step, which makes it infeasible to synthesize robust plans based on relative delays.
By introducing global time points, we decouple the controller decisions from the nondeterministic timing behavior of the environment.  
The synthesized plan specifies \emph{when} transitions must occur in absolute time, rather than \emph{how long} to wait locally.  
This enables the computation of plans that are robust against all admissible system and environment behaviors.

One question for discussion is how scalable this approach is compared to synthesizing a full reactive strategy. Currently, we can only offer an intuitive answer to this question. 
A naive, limited SMT encoding of a fully reactive strategy like we can synthesize  with UPPAAL \cite{behrmann2006uppaal, behrmann2007uppaal, Uppaalstratego} would require alternating decisions regarding successive actions from the environment and the system. For this further nesting of quantifiers (\ie{} $\exists \, \forall \, \exists \cdots$) would be required across the entire number of steps $k$. This would correspond to a nested depth of approximately $2k$, which would entail enormous computational complexity. Our approach fundamentally reduces the nesting to a depth of two, so intuitively, it scales better than computing a full reactive strategy.
This differs from the approach implemented in UPPAAL TIGA, which is based on symbolic, zone-based game theory. Our comparison does not claim that UPPAAL TIGA directly implements such an alternating quantifier structure. Rather, our approach tackles a different design aspect: the synthesis of fixed schedules in global time rather than state-dependent strategies. Another conceptual benefit of our approach is the simpler integration of SMT theories. A direct runtime evaluation with \cite{kroger2023updates} is left as a topic for future work. Beside that an interesting question arises: Is it possible to optimize the game-based extension for BMC to achieve a comparable level of scalability, similar to the optimized synthesis of a reactive strategy for timed games in UPPAAL?

%% file: sections/sectrelatedwork.tex
\section{Related Work}\label{sec:relatedwork}
Approaches to update scheduling and strategy synthesis can be broadly divided into two directions: (i) methods that determine safe system updates under given specifications, and (ii) algorithmic techniques and tools for synthesizing strategies, in particular based on timed games and SMT-based BMC.

\textit{\textbf{Update Scheduling:}} Holthusen et al.~\cite{holthusen2016using} propose a contract-based methodology for performing software updates in safety-critical embedded systems. Their approach replaces traditional validation with formal analyses that determine whether an update can be applied to a given configuration. A negotiation controller iteratively explores the configuration space and refines it by adding constraints whenever violations occur, until a valid configuration is found or infeasibility is established.
However, the contract model does not enforce a strict separation between assumptions and guarantees and thus implicitly relies on cooperative system behavior. In contrast, our approach models the interaction as a reactive game and ensures correctness under all possible environment behaviors via universal quantification.
Finkbeiner and Schewe~\cite{finkbeiner2013bounded} introduce bounded synthesis, which reduces strategy synthesis from LTL specifications to an SMT problem by bounding the number of states. While this approach shares the idea of SMT-based synthesis, it focuses on discrete systems. In contrast, our work considers timed systems and synthesizes global time points that define robust schedules.
Göttmann et al.~\cite{gottmann2024cost} model runtime reconfiguration as a stochastic priced timed game and synthesize strategies using \textit{UPPAAL Stratego} \cite{Uppaalstratego}. Their approach focuses on quantitative optimization under stochastic behavior, whereas we target qualitative guarantees in adversarial settings and derive schedules that are robust for all executions.

\textit{\textbf{Tools for Synthesis:}} Timed game solving is supported by tools such as \textit{UPPAAL TIGA}~\cite{behrmann2006uppaal}, which synthesizes strategies for reachability and safety objectives based on controllable and uncontrollable transitions. However, its expressiveness is limited when combining complex objectives.
\textit{UPPAAL Stratego}~\cite{Uppaalstratego} extends this approach to stochastic and priced timed games, enabling optimization and evaluation of strategies. While effective for quantitative analysis, it is less suited for symbolic reasoning with universally quantified adversarial behaviors.

SMT solvers provide a flexible basis for BMC-based synthesis. Tools such as \textit{ISAT3}~\cite{scheibler2014implication, scheibler2014isat3} support bounded model checking but are restricted to existential reasoning. Other solvers, including \textit{Boolector}~\cite{brummayer2009boolector} and \textit{Yices}~\cite{dutertre2006yices}, do not support quantifiers. More expressive solvers such as \textit{cvc5}~\cite{barbosa2022cvc5} and \textit{MathSAT5}~\cite{cimatti2013mathsat5} support quantified formulas but often struggle with nested and alternating quantifiers.

In contrast, our approach explicitly targets SMT-based synthesis with alternating quantifiers arising from the game-theoretic formulation and adapts the encoding to handle timing constraints and robustness requirements.

%% file: sections/sectconclusion.tex
\section{Conclusion}\label{sec:conclusion}
In this paper, we present a method for synthesizing robust plans for scheduling updates using game semantics in combination with BMC. By modeling the update process as a timed game between the update and the system and quantifier alternation, we can synthesize global time points for update steps independent of the system behavior.
Unlike reactive strategies, in which actions of the update depend directly on the actions of the system, a robust plan provides a fixed, guaranteed schedule for performing updates. We demonstrate our synthesis using the running example of a \textit{Trajectory Planner} and synthesize a robust update plan with the SMT solver Z3 \cite{de2008z3}.
These results indicates that robust plans are a promising approach for ensuring safe and predictable system updates. In the future, we plan to conduct a detailed scalability analysis and compare it with reactive strategy synthesis approaches in order to identify any additional advantages.

%% file: SynthesizingUpdateScheduleBMC.bbl
\begin{thebibliography}{10}
\providecommand{\bibitemdeclare}[2]{}
\providecommand{\surnamestart}{}
\providecommand{\surnameend}{}
\providecommand{\urlprefix}{Available at }
\providecommand{\url}[1]{\texttt{#1}}
\providecommand{\href}[2]{\texttt{#2}}
\providecommand{\urlalt}[2]{\href{#1}{#2}}
\providecommand{\doi}[1]{doi:\urlalt{https://doi.org/#1}{#1}}
\providecommand{\eprint}[1]{arXiv:\urlalt{https://arxiv.org/abs/#1}{#1}}
\providecommand{\bibinfo}[2]{#2}

\bibitemdeclare{article}{alur1994theory}
\bibitem{alur1994theory}
\bibinfo{author}{Rajeev \surnamestart Alur\surnameend} \&
  \bibinfo{author}{David~L \surnamestart Dill\surnameend}
  (\bibinfo{year}{1994}): \emph{\bibinfo{title}{A theory of timed automata}}.
\newblock {\slshape \bibinfo{journal}{Theoretical computer science}}
  \bibinfo{volume}{126}(\bibinfo{number}{2}), pp. \bibinfo{pages}{183--235},
  \doi{10.1016/0304-3975(94)90010-8}.

\bibitemdeclare{inproceedings}{barbosa2022cvc5}
\bibitem{barbosa2022cvc5}
\bibinfo{author}{Haniel \surnamestart Barbosa\surnameend},
  \bibinfo{author}{Clark \surnamestart Barrett\surnameend},
  \bibinfo{author}{Martin \surnamestart Brain\surnameend},
  \bibinfo{author}{Gereon \surnamestart Kremer\surnameend},
  \bibinfo{author}{Hanna \surnamestart Lachnitt\surnameend},
  \bibinfo{author}{Makai \surnamestart Mann\surnameend},
  \bibinfo{author}{Abdalrhman \surnamestart Mohamed\surnameend},
  \bibinfo{author}{Mudathir \surnamestart Mohamed\surnameend},
  \bibinfo{author}{Aina \surnamestart Niemetz\surnameend},
  \bibinfo{author}{Andres \surnamestart N{\"o}tzli\surnameend} et~al.
  (\bibinfo{year}{2022}): \emph{\bibinfo{title}{cvc5: A versatile and
  industrial-strength SMT solver}}.
\newblock In: {\slshape \bibinfo{booktitle}{International Conference on Tools
  and Algorithms for the Construction and Analysis of Systems}},
  \bibinfo{organization}{Springer}, pp. \bibinfo{pages}{415--442},
  \doi{10.1007/978-3-030-99524-9_24}.

\bibitemdeclare{article}{behrmann2007uppaal}
\bibitem{behrmann2007uppaal}
\bibinfo{author}{Gerd \surnamestart Behrmann\surnameend},
  \bibinfo{author}{Agnes \surnamestart Cougnard\surnameend},
  \bibinfo{author}{Alexandre \surnamestart David\surnameend},
  \bibinfo{author}{Emmanuel \surnamestart Fleury\surnameend},
  \bibinfo{author}{Kim~G \surnamestart Larsen\surnameend} \&
  \bibinfo{author}{Didier \surnamestart Lime\surnameend}
  (\bibinfo{year}{2007}): \emph{\bibinfo{title}{Uppaal tiga user-manual}}.
\newblock {\slshape \bibinfo{journal}{Aalborg University}}.

\bibitemdeclare{inproceedings}{behrmann2006uppaal}
\bibitem{behrmann2006uppaal}
\bibinfo{author}{Gerd \surnamestart Behrmann\surnameend},
  \bibinfo{author}{Agnes \surnamestart Cougnard\surnameend},
  \bibinfo{author}{Alexandre \surnamestart David\surnameend},
  \bibinfo{author}{Emmanuel \surnamestart Fleury\surnameend},
  \bibinfo{author}{Kim~Guldstrand \surnamestart Larsen\surnameend} \&
  \bibinfo{author}{Didier \surnamestart Lime\surnameend}
  (\bibinfo{year}{2006}): \emph{\bibinfo{title}{UPPAAL-Tiga: Timed games for
  everyone}}.
\newblock In: {\slshape \bibinfo{booktitle}{Nordic Workshop on Programming
  Theory (NWPT'06)}}.

\bibitemdeclare{incollection}{Bloem2018}
\bibitem{Bloem2018}
\bibinfo{author}{Roderick \surnamestart Bloem\surnameend},
  \bibinfo{author}{Krishnendu \surnamestart Chatterjee\surnameend} \&
  \bibinfo{author}{Barbara \surnamestart Jobstmann\surnameend}
  (\bibinfo{year}{2018}): \emph{\bibinfo{title}{Graph Games and Reactive
  Synthesis}}.
\newblock In \bibinfo{editor}{Edmund~M. \surnamestart Clarke\surnameend},
  \bibinfo{editor}{Thomas~A. \surnamestart Henzinger\surnameend},
  \bibinfo{editor}{Helmut \surnamestart Veith\surnameend} \&
  \bibinfo{editor}{Roderick \surnamestart Bloem\surnameend}, editors: {\slshape
  \bibinfo{booktitle}{Handbook of Model Checking}},
  \bibinfo{publisher}{Springer International Publishing},
  \bibinfo{address}{Cham}, pp. \bibinfo{pages}{921--962},
  \doi{10.1007/978-3-319-10575-8_27}.

\bibitemdeclare{inproceedings}{brummayer2009boolector}
\bibitem{brummayer2009boolector}
\bibinfo{author}{Robert \surnamestart Brummayer\surnameend} \&
  \bibinfo{author}{Armin \surnamestart Biere\surnameend}
  (\bibinfo{year}{2009}): \emph{\bibinfo{title}{Boolector: An efficient SMT
  solver for bit-vectors and arrays}}.
\newblock In: {\slshape \bibinfo{booktitle}{International Conference on Tools
  and Algorithms for the Construction and Analysis of Systems}},
  \bibinfo{organization}{Springer}, pp. \bibinfo{pages}{174--177},
  \doi{10.1007/978-3-642-00768-2_16}.

\bibitemdeclare{inproceedings}{cimatti2013mathsat5}
\bibitem{cimatti2013mathsat5}
\bibinfo{author}{Alessandro \surnamestart Cimatti\surnameend},
  \bibinfo{author}{Alberto \surnamestart Griggio\surnameend},
  \bibinfo{author}{Bastiaan~Joost \surnamestart Schaafsma\surnameend} \&
  \bibinfo{author}{Roberto \surnamestart Sebastiani\surnameend}
  (\bibinfo{year}{2013}): \emph{\bibinfo{title}{The mathsat5 smt solver}}.
\newblock In: {\slshape \bibinfo{booktitle}{International Conference on Tools
  and Algorithms for the Construction and Analysis of Systems}},
  \bibinfo{organization}{Springer}, pp. \bibinfo{pages}{93--107},
  \doi{10.1007/978-3-642-36742-7_7}.

\bibitemdeclare{inproceedings}{Uppaalstratego}
\bibitem{Uppaalstratego}
\bibinfo{author}{Alexandre \surnamestart David\surnameend},
  \bibinfo{author}{Peter~G. \surnamestart Jensen\surnameend},
  \bibinfo{author}{Kim~Guldstrand \surnamestart Larsen\surnameend},
  \bibinfo{author}{Marius \surnamestart Miku{\v c}ionis\surnameend} \&
  \bibinfo{author}{Jakob~H. \surnamestart Taankvist\surnameend}
  (\bibinfo{year}{2015}): \emph{\bibinfo{title}{UPPAAL Stratego}}.
\newblock In \bibinfo{editor}{Christel \surnamestart Baier\surnameend} \&
  \bibinfo{editor}{Cesare \surnamestart Tinelli\surnameend}, editors: {\slshape
  \bibinfo{booktitle}{Tools and Algorithms for the Construction and Analysis of
  Systems}}, \bibinfo{publisher}{Springer Berlin Heidelberg},
  \bibinfo{address}{Berlin, Heidelberg}, pp. \bibinfo{pages}{206--211},
  \doi{10.1007/978-3-662-46681-0_16}.

\bibitemdeclare{inproceedings}{de2008z3}
\bibitem{de2008z3}
\bibinfo{author}{Leonardo \surnamestart De~Moura\surnameend} \&
  \bibinfo{author}{Nikolaj \surnamestart Bj{\o}rner\surnameend}
  (\bibinfo{year}{2008}): \emph{\bibinfo{title}{Z3: An efficient SMT solver}}.
\newblock In: {\slshape \bibinfo{booktitle}{International conference on Tools
  and Algorithms for the Construction and Analysis of Systems}},
  \bibinfo{organization}{Springer}, pp. \bibinfo{pages}{337--340},
  \doi{10.1007/978-3-540-78800-3_24}.

\bibitemdeclare{article}{dutertre2006yices}
\bibitem{dutertre2006yices}
\bibinfo{author}{Bruno \surnamestart Dutertre\surnameend} \&
  \bibinfo{author}{Leonardo \surnamestart De~Moura\surnameend}
  (\bibinfo{year}{2006}): \emph{\bibinfo{title}{The yices smt solver}}.
\newblock {\slshape \bibinfo{journal}{Tool paper at
  http://yices.csl.sri.com/tool-paper.pdf}}
  \bibinfo{volume}{2}(\bibinfo{number}{2}), pp. \bibinfo{pages}{1--2}.

\bibitemdeclare{article}{finkbeiner2013bounded}
\bibitem{finkbeiner2013bounded}
\bibinfo{author}{Bernd \surnamestart Finkbeiner\surnameend} \&
  \bibinfo{author}{Sven \surnamestart Schewe\surnameend}
  (\bibinfo{year}{2013}): \emph{\bibinfo{title}{Bounded synthesis}}.
\newblock {\slshape \bibinfo{journal}{International Journal on Software Tools
  for Technology Transfer}} \bibinfo{volume}{15}(\bibinfo{number}{5}), pp.
  \bibinfo{pages}{519--539}, \doi{10.1007/s10009-012-0228-z}.

\bibitemdeclare{article}{gottmann2024cost}
\bibitem{gottmann2024cost}
\bibinfo{author}{Hendrik \surnamestart G{\"o}ttmann\surnameend},
  \bibinfo{author}{Birte \surnamestart Caesar\surnameend},
  \bibinfo{author}{Lasse \surnamestart Beers\surnameend},
  \bibinfo{author}{Malte \surnamestart Lochau\surnameend},
  \bibinfo{author}{Andy \surnamestart Sch{\"u}rr\surnameend} \&
  \bibinfo{author}{Alexander \surnamestart Fay\surnameend}
  (\bibinfo{year}{2024}): \emph{\bibinfo{title}{Cost-sensitive precomputation
  of real-time-aware reconfiguration strategies based on stochastic priced
  timed games}}.
\newblock {\slshape \bibinfo{journal}{Software and Systems Modeling}}, pp.
  \bibinfo{pages}{1--31}, \doi{10.1007/s10270-024-01195-9}.

\bibitemdeclare{inproceedings}{guissouma2018empirical}
\bibitem{guissouma2018empirical}
\bibinfo{author}{Houssem \surnamestart Guissouma\surnameend},
  \bibinfo{author}{Heiko \surnamestart Klare\surnameend}, \bibinfo{author}{Eric
  \surnamestart Sax\surnameend} \& \bibinfo{author}{Erik \surnamestart
  Burger\surnameend} (\bibinfo{year}{2018}): \emph{\bibinfo{title}{An empirical
  study on the current and future challenges of automotive software release and
  configuration management}}.
\newblock In: {\slshape \bibinfo{booktitle}{2018 44th Euromicro Conference on
  Software Engineering and Advanced Applications (SEAA)}},
  \bibinfo{organization}{IEEE}, pp. \bibinfo{pages}{298--305},
  \doi{10.1109/SEAA.2018.00056}.

\bibitemdeclare{article}{holthusen2016using}
\bibitem{holthusen2016using}
\bibinfo{author}{S{\"o}nke \surnamestart Holthusen\surnameend},
  \bibinfo{author}{Sophie \surnamestart Quinton\surnameend},
  \bibinfo{author}{Ina \surnamestart Schaefer\surnameend},
  \bibinfo{author}{Johannes \surnamestart Schlatow\surnameend} \&
  \bibinfo{author}{Martin \surnamestart Wegner\surnameend}
  (\bibinfo{year}{2016}): \emph{\bibinfo{title}{Using multi-viewpoint contracts
  for negotiation of embedded software updates}}.
\newblock {\slshape \bibinfo{journal}{arXiv preprint arXiv:1606.00504}},
  \doi{10.4204/EPTCS.208.3}.

\bibitemdeclare{inproceedings}{kroger2023updates}
\bibitem{kroger2023updates}
\bibinfo{author}{Janis \surnamestart Kr{\"o}ger\surnameend} \&
  \bibinfo{author}{Martin \surnamestart Fr{\"a}nzle\surnameend}
  (\bibinfo{year}{2023}): \emph{\bibinfo{title}{Updates at Runtime for Cyber
  Physical Systems. A Game Theoretic Approach}}.
\newblock In: {\slshape \bibinfo{booktitle}{Software Engineering 2023
  Workshops}}, \bibinfo{organization}{Gesellschaft f{\"u}r Informatik eV}, pp.
  \bibinfo{pages}{54--65}, \doi{10.18420/se2023-ws-08}.

\bibitemdeclare{inproceedings}{kroger2025ensuring}
\bibitem{kroger2025ensuring}
\bibinfo{author}{Janis \surnamestart Kr{\"o}ger\surnameend},
  \bibinfo{author}{Ingo \surnamestart Stierand\surnameend} \&
  \bibinfo{author}{Martin \surnamestart Fr{\"a}nzle\surnameend}
  (\bibinfo{year}{2025}): \emph{\bibinfo{title}{Ensuring Integration Conditions
  During the Update of Cyber-Physical Systems at Runtime}}.
\newblock In: {\slshape \bibinfo{booktitle}{International Conference on Formal
  Methods for Industrial Critical Systems}}, \bibinfo{organization}{Springer},
  pp. \bibinfo{pages}{203--221}, \doi{10.1007/978-3-032-00942-5_11}.

\bibitemdeclare{book}{maschler2020game}
\bibitem{maschler2020game}
\bibinfo{author}{Michael \surnamestart Maschler\surnameend},
  \bibinfo{author}{Shmuel \surnamestart Zamir\surnameend} \&
  \bibinfo{author}{Eilon \surnamestart Solan\surnameend}
  (\bibinfo{year}{2020}): \emph{\bibinfo{title}{Game theory}}.
\newblock \bibinfo{publisher}{Cambridge University Press},
  \doi{10.1017/9781108636049}.

\bibitemdeclare{inproceedings}{rakow2022roles}
\bibitem{rakow2022roles}
\bibinfo{author}{Astrid \surnamestart Rakow\surnameend} \&
  \bibinfo{author}{Janis \surnamestart Kr{\"o}ger\surnameend}
  (\bibinfo{year}{2022}): \emph{\bibinfo{title}{Roles and Responsibilities for
  a Predictable Update Process--A Position Paper}}.
\newblock In: {\slshape \bibinfo{booktitle}{International Conference on
  Verification and Evaluation of Computer and Communication Systems}},
  \bibinfo{organization}{Springer}, pp. \bibinfo{pages}{17--26},
  \doi{10.1007/978-3-030-98850-0_2}.

\bibitemdeclare{manual}{scheibler2014isat3}
\bibitem{scheibler2014isat3}
\bibinfo{author}{Karsten \surnamestart Scheibler\surnameend}
  (\bibinfo{year}{2014}): \emph{\bibinfo{title}{iSAT3 Manual}}.

\bibitemdeclare{inproceedings}{scheibler2014implication}
\bibitem{scheibler2014implication}
\bibinfo{author}{Karsten \surnamestart Scheibler\surnameend},
  \bibinfo{author}{Bernd \surnamestart Becker\surnameend},
  \bibinfo{author}{J~\surnamestart Ruf\surnameend},
  \bibinfo{author}{D~\surnamestart Allmendinger\surnameend} \&
  \bibinfo{author}{M~\surnamestart Michel\surnameend} (\bibinfo{year}{2014}):
  \emph{\bibinfo{title}{Implication Graph Compression inside the SMT Solver
  iSAT3}}.
\newblock In: {\slshape \bibinfo{booktitle}{MBMV}}, pp.
  \bibinfo{pages}{25--36}.

\end{thebibliography}
